\documentclass [12pt]{article}
\usepackage{graphicx}

\begin{document}

{\bf Does an irreversible chemical cycle support equilibrium?}


\vskip .1in

Kinshuk Banerjee and Kamal Bhattacharyya\footnote{Corresponding 
author; e-mail: pchemkb@gmail.com}

\vskip .05in
{\it Department of Chemistry, University of Calcutta, 
92 A.P.C. Road,

Kolkata 700 009, India.}

\begin{abstract}

The impossibility of attaining equilibrium for cyclic chemical reaction  networks with irreversible steps is apparently due to a 
divergent entropy production rate. A deeper reason seems to be the 
violation of the detailed balance condition. 
In this work, 
we discuss how the standard theoretical framework can be 
adapted to include irreversible cycles, avoiding the divergence. 
With properly redefined force terms, such systems are also 
seen to reach and sustain equilibria that are characterized by the 
vanishing 
of the entropy production rate, though detailed 
balance is not maintained. 
Equivalence of the present formulation with Onsager's original 
prescription 
is established for both reversible and irreversible 
cycles, with a few adjustments in the latter case. 
Further justification of the attainment of true equilibrium 
is provided with the help of 
the minimum entropy production principle. 
All the results are generalized for an irreversible cycle 
comprising of $N$ number of species. 

\end{abstract}

\section{Introduction}

Chemical reactions play a very important and interesting 
part in the theory and applications of non-equilibrium 
thermodynamics since inception \cite{ons1,ons2,denb,groot,groot1}. 
The irreversibility of processes in real systems 
\cite{prigogn1,prigogn2,prigogn3,konde} all around us and 
inside our bodies 
are almost always connected to chemical reactions \cite{Hill1}. 
Thus, as was the case with equilibrium thermodynamics, major 
attention is paid towards chemical systems throughout the 
development of this discipline \cite{gasprd,gasprd1,vel,hqian,Niven}. 
The key concept of entropy production rate (EPR) 
\cite{Schnak,Min,Zhang,Jiang,Hqian1,Hqian2,KBan}
to describe irreversible processes was connected with 
the reaction affinity since the very early days of irreversible 
thermodynamics by de Donder \cite{deDon}. 
In his seminal paper,  
Onsager \cite{ons1} introduced the reciprocal relations 
by considering a reversible triangular 
reaction network. 
He 
noted, however, that `detailed balance' acts as an 
{\it additional restriction} 
to describe chemical equilibrium, apart from the 
second law of thermodynamics. 
Since then, this `additional restriction' of detailed 
balance became a rule of thumb, 
requiring each reaction to contain elementary steps in 
forward and reverse directions. Both these steps need 
to be considered in the formulation of EPR 
\cite{nicol,sevk,espo,jarz2011,seifert}.

The above theoretical scheme fails to account for any 
equilibria in 
irreversible chemical cycles 
because setting of `backward' rate constants equal to zero 
leads to divergence of the EPR. This is certainly 
unphysical. 
Unfortunately however, such an outcome 
has been used as an argument to eliminate the possibility 
of equilibrium being sustained by reaction 
cycles with irreversible steps, although 
the corresponding kinetic equations give fully 
consistent results. 
Recently, a few studies have addressed the issue. 
One bypass route is to 
coarse grain the system evolution at regular time intervals 
so that effective transition rates can be defined \cite{ben}. 
Another proposition is to perform an experiment over such a 
time span that 
the backward step, although present, has no chance to occur \cite{Zer}. 
In this latter work \cite{Zer}, the authors rigorously derived a formula 
for EPR for irreversible transitions at the microscopic 
level, showing logarithmic dependence on the time span chosen 
and argued that the divergence of EPR for 
such processes is a theoretical artifact. 
This technique, however, is based on how accurately the 
setting of the finite time span approximates a real 
irreversible process, so as to avoid the divergence of EPR. 
Understandably, the 
backward rate constant is considered to be {\it very small}, but 
{\it not exactly zero}.

Having stated the background, 
here we study the EPR in cyclic chemical reaction networks 
with {\it exactly} irreversible steps, {\it i.e.,} 
we set 
all the backward rate constants {\it equal to zero}. 
We show that the standard 
formulation of EPR in terms of the 
flux-force relations of individual reactions \cite{groot,gasprd1,vel} 
can be applied to these systems as well, but 
only after proper {\it modifications}. 
Starting with the simplest example of a triangular network, the 
EPR is shown to vanish at the steady state of the 
irreversible cycle, justifying that the system reaches a true 
thermodynamic equilibrium. 
We further import the minimum entropy production principle 
\cite{ross1,ross2} that unequivocally ascertains 
the nature of a steady state. 
Finally, we generalize all the findings to an irreversible 
cycle containing $N$ number of species to witness 
similar features.

\section{Problem with irreversible cycles}

\begin{figure}[tbh]
\centering
\rotatebox{270}{
\includegraphics[width=6cm,keepaspectratio]{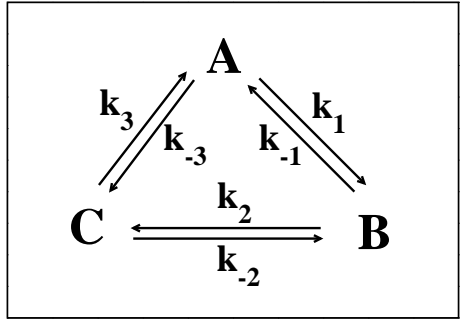}}
\caption{Schematic diagram of the ABC cyclic reversible reaction network}
\label{fig1}
\end{figure}

We start the discussion with a reversible cycle consisting 
of three species, A, B, C as shown in Fig.\ref{fig1}. 
The time-dependent concentrations of species A, B, C, 
are denoted by $a(t), b(t), c(t)$, respectively. 
The rate equations are given as 
\begin{equation}
\dot{a}=-(k_1+k_{-3})a(t)+k_{-1}b(t)+k_3c(t),
\label{abc1n}
\end{equation}
\begin{equation}
\dot{b}=k_1a(t)-(k_{-1}+k_2)b(t)+k_{-2}c(t),
\label{abc2n}
\end{equation}
\begin{equation}
\dot{c}=k_{-3}a(t)+k_2b(t)-(k_{-2}+k_3)c(t),
\label{abc3n}
\end{equation}
with $\dot{a}+\dot{b}+\dot{c}=0.$

The EPR $\sigma(t)$ of the network is expressed in terms 
of fluxes $J_i$ and the corresponding forces $X_i$ as \cite{groot}
\begin{equation}
\sigma(t)=\frac{1}{T}\sum_{i=1}^{3} J_i(t) X_i(t).
\label{epr}
\end{equation}
The fluxes are defined as \cite{groot,nicol,espo}: 
\begin{equation}
J_1(t)=k_{-1}b(t)-k_1 a(t),
\label{j1}
\end{equation}
\begin{equation}
J_2(t)=k_{-2}c(t)-k_2 b(t),
\label{j2}
\end{equation}
\begin{equation}
J_3(t)=k_{-3}a(t)-k_3 c(t).
\label{j3}
\end{equation}
The corresponding forces are 
\begin{equation}
X_1(t)=\mu_B-\mu_A=T{\rm ln}\frac{k_{-1}b(t)}{k_1 a(t)},
\label{x1}
\end{equation}
\begin{equation}
X_2(t)=\mu_C-\mu_B=T{\rm ln}\frac{k_{-2}c(t)}{k_2 b(t)},
\label{x2}
\end{equation}
\begin{equation}
X_3(t)=\mu_A-\mu_C=T{\rm ln}\frac{k_{-3}a(t)}{k_3 c(t)}
\label{x3}
\end{equation}
Here and throughout the paper, we have set the Boltzmann constant 
$k_B=1$. 
The condition of detailed balance \cite{denb,groot} requires 
the fluxes of each individual reaction to vanish at steady state, 
{\it i.e.}, 
\begin{equation}
J_1^s=J_2^s=J_3^s=0.
\label{fluxeq}
\end{equation} 
When this condition is satisfied, the reaction system reaches 
equilibrium.

Note that, if the steps are irreversible, we must set $k_{-1}=
k_{-2}=k_{-3}=0$. 
Then, the forces given by right sides in Eqs (\ref{x1}-\ref{x3}), 
and hence the EPR, diverge. 
This necessitates a different approach where, obviously, the forces 
need to be redefined to avoid any disaster. 
So, the basic problem is to have a 
divergence-free expression for the EPR in an irreversible cycle. 

However, before proceeding further, we make some 
comments on the prevailing notion that 
equilibrium can not be maintained by an irreversible cycle 
due to (i) divergent nature of EPR and (ii) violation of the 
detailed balance condition. 
Indeed, Onsager himself considered detailed balance as an 
{\it additional assumption}. He wrote \cite{ons1}:
``..., however, the chemists are accustomed to impose a very 
interesting additional restriction, namely: when the equilibrium 
is reached each individual reaction must balance itself.''
This point is also discussed at lenght by Denbigh, clearly 
stating that the above system {\it can} reach equilibrium following 
the laws of thermodynamics without requiring the 
condition of detailed balance \cite{denb}. 
Indeed, if we investigate the role of the 
backward step in each individual reaction, we see that 
it provides a pathway that produces the opposite 
effect of that due to the forward step on the concentrations of the 
species involved. Now, the beauty of the cyclic 
network is that, even with irreversible steps, there 
exists a `feedback' for each species, although not 
in the sense of detailed balance. 
Thus, all the species in 
the irreversible cycle have finite, non-zero 
concentrations and well-defined chemical potentials 
during the time evolution as well as in the long-time limit. 
In our opinion, this provides enough justification 
to search for a consistent non-equilibrium thermodynamic 
description of such systems. 
In this context, we may also mention the work of Xiao {\it et al.} 
\cite{Xiao}
on the entropy production in a Brusselator model 
with irreversible steps where the state changes in the 
population space become reversible.


\section{The remedy}

\begin{figure}[tbh]
\centering
\rotatebox{270}{
\includegraphics[width=6cm,keepaspectratio]{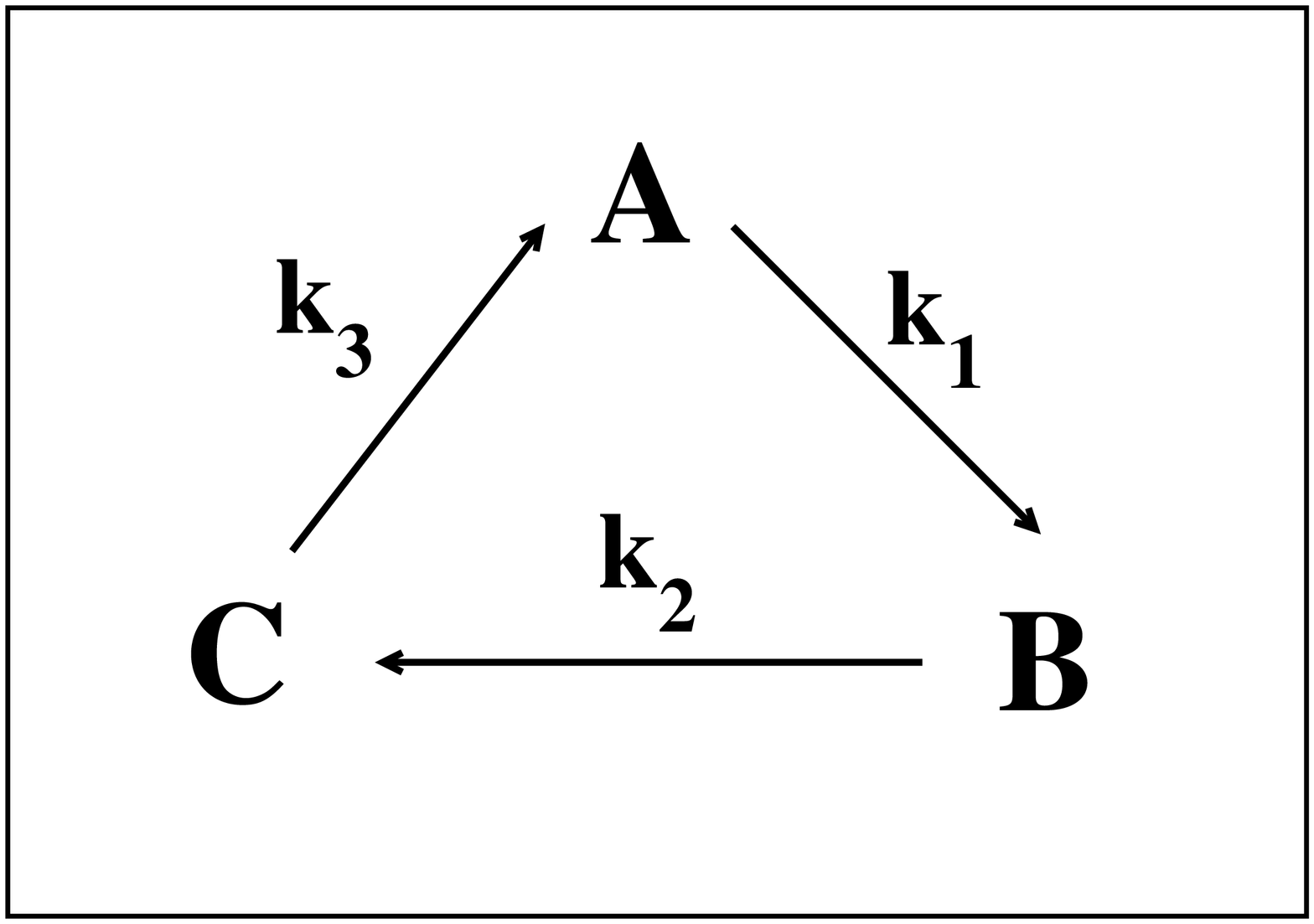}}
\caption{Schematic diagram of the ABC cyclic irreversible reaction network}
\label{fig2}
\end{figure}

The kinetic scheme of the ABC irreversible cycle 
is shown in Fig.\ref{fig2}. 
At first, we determine the equilibrium concentrations of 
the species. 
They are obtained from Eqs (\ref{abc1n})-(\ref{abc3n}) by 
setting $\dot{a}=\dot{b}=\dot{c}=0$ with 
$k_{-1}=k_{-2}=k_{-3}=0$, as 
\begin{equation}
a^e=k_2k_3/K,
\label{air}
\end{equation}
\begin{equation}
b^e=k_1k_3/K,
\label{bir}
\end{equation}
\begin{equation}
c^e=k_1k_2/K,
\label{cir}
\end{equation}
with $K=k_1k_2+k_2k_3+k_1k_3.$ 
By equilibrium, we mean the true thermodynamic equilibrium only with 
\begin{equation}
\mu_A^e=\mu_B^e=\mu_C^e.
\label{eqlbm}
\end{equation}
Then, using the above equilibrium concentrations and 
Eq.(\ref{eqlbm}), 
we obtain the expressions of the forces in the irreversible 
cycle as
\begin{equation}
X_1=(\mu_B-\mu_A)=\Delta \mu^0_{BA}+T{\rm ln}\frac{b}{a}
=-T{\rm ln}\frac{b^e}{a^e}+ T{\rm ln}\frac{b}{a}=
T{\rm ln}\frac{k_2b}{k_1a}.
\label{muca}
\end{equation}
Similarly,
\begin{equation}
X_2=(\mu_C-\mu_B)=T{\rm ln}\frac{k_3c}{k_2b},
\label{mucb}
\end{equation}
\begin{equation}
X_3=(\mu_A-\mu_C)=T{\rm ln}\frac{k_1a}{k_3c}.
\label{mucc}
\end{equation}
The above equations are 
naturally adjusted to avoid the divergence experienced 
in case of their counterparts in the reversible cycle, given in 
Eqs (\ref{x1})-(\ref{x3}). 
In the derivation of Eqs (\ref{muca})-(\ref{mucc}), the 
major modification is in the definition of equilibrium constants 
appearing in the standard-state chemical potential differences. 
Conventionally, the equilibrium constant is 
taken equal to the ratio of forward and reverse rate constants. 
When the reactions are irreversible, there 
are no reverse rate constants and hence, one needs to 
generalize the concept. 
In Eqs (\ref{muca})-(\ref{mucc}), the ratio of the 
concentrations of the species, constant at equilibrium, 
plays the role of an equilibrium constant. 
In this context, we mention that 
the equality of equilibrium constant to the ratio of forward and 
backward rate constants is often misunderstood 
as a consequence of thermodynamics. 
As stated by Denbigh \cite{denb}:``The point made by 
Onsager is that this equality can 
be proved from the second law only in the special case where 
there is a single independent reaction taking place in the 
system.'' Thus, for multiple reactions, the conventional 
definition of equilibrium constant requires 
detailed balance to hold at equilibrium. 
This stands as an 
{\it extra} principle not contained in thermodynamics. 
Hence, in an irreversible cycle where detailed 
balance is violated, 
it is not surprising that the equilibrium 
constant should be modified accordingly. 
Now, from Eqs (\ref{j1})-(\ref{j3}), the fluxes in the irreversible 
cycle become 
\begin{equation}
J_1(t)=-k_1 a(t),
\label{j1ir}
\end{equation}
\begin{equation}
J_2(t)=-k_2 b(t),
\label{j2ir}
\end{equation}
\begin{equation}
J_3(t)=-k_3 c(t).
\label{j3ir}
\end{equation}
Then, from Eq.(\ref{epr}), we obtain the EPR of the irreversible cycle as
\begin{equation}
\sigma(t)=k_1a(t){\rm ln}\frac{k_1a(t)}{k_2b(t)}
+k_2b(t){\rm ln}\frac{k_2b(t)}{k_3c(t)}+
k_3c(t){\rm ln}\frac{k_3c(t)}{k_1a(t)},
\label{eprir}
\end{equation}
which obviously remains finite. 
Still, the acid test for any expression of EPR is its positivity. 
In the next paragraph, we will show that $\sigma(t)$, as given 
in Eq.(\ref{eprir}), is always positive and becomes 
zero at equilibrium. 

We define 
\begin{equation}
k_1a(t)+k_2b(t)+k_3c(t)=N(t)>0.
\label{prf1}
\end{equation}
Then dividing Eq.(\ref{eprir}) by $N(t)$, we get
\begin{equation}
\frac{1}{N(t)}\sigma(t)=x{\rm ln}\frac{x}{y}+y{\rm ln}\frac{y}{z}+
z{\rm ln}\frac{z}{x},
\label{prf2}
\end{equation}
where 
$$x=k_1a(t)/N(t),$$ 
$$y=k_2b(t)/N(t),$$ 
$$z=k_3c(t)/N(t),$$ 
with $$x+y+z=1,$$  
from Eq.(\ref{prf1}) . 
Now, we define two normalized probability distributions, 
${\bf P}(=\{p_i\},i=1,2,3)$ and ${\bf Q}(=\{q_i\},i=1,2,3)$, 
with 
\begin{equation}
p_1=x,\,p_2=y,\,p_3=z,
\label{prb1}
\end{equation}
and
\begin{equation}
q_1=y,\,q_2=z,\,q_3=x.
\label{prb2}
\end{equation}
Then we can rewrite Eq.(\ref{prf2}) in a more revealing 
form of the Kullback-Leibler distance of 
two normalized distributions \cite{KL} and using the 
positivity of the latter \cite{CovThom}, the proof is complete by 
virtue of the expression 
\begin{equation}
\sigma(t)=N(t)\sum_{i=1}^3 p_i{\rm ln}\frac{p_i}{q_i}\ge 0.
\label{prf3}
\end{equation}
The equality in Eq.(\ref{prf3}) will be valid for 
$p_i=q_i, \,\forall i.$ This implies $x=y=z$, which in 
turn leads to $k_1a=k_2b=k_3c.$ It is easy to see 
from Eqs (\ref{abc1n})-(\ref{abc3n}) with 
$k_{-1}=k_{-2}=k_{-3}=0$
that this condition will be satisfied 
when $\dot{a}=\dot{b}=\dot{c}=0.$ 
So EPR becomes zero as the concentrations of all the species 
become fixed. 
Then we can say that the irreversible cycle reaches equilibrium, 
characterized by the vanishing of EPR, as desired. 
Hence, the expression of EPR for the irreversible cycle 
in Eq.(\ref{eprir}) satisfies all the basic requirements 
to be thermodynamically consistent and physically meaningful.

For further support of the result that such an irreversible 
cycle reaches equilibrium as obtained above, we explore the 
minimum entropy production principle 
(MEPP) \cite{groot1,prigogn2}. 
To this end, we temporarily call the equilibrium 
to be a steady state.  
MEPP tells that EPR has its minimum at a steady state 
that lies close enough to equilibrium  
with approximately linear relation between fluxes and forces \cite{prigogn2}. 
However, recently 
it has been rigorously shown by Ross and coauthors \cite{ross1,ross2}, 
that this principle 
is true if and only if the steady state is the state 
of thermodynamic equilibrium. 
Here we investigate this fact by using the expression 
EPR of the irreversible cycle. 
From Eq.(\ref{eprir}), we get
$$\left(\frac{\partial\sigma'}{\partial a}\right)=
k_1{\rm ln}\frac{k_1a}{k_2b}+\frac{k_1a-k_3c}{a}.$$
At steady state, $\dot{a}=\dot{b}=\dot{c}=0$ and 
from Eqs (\ref{abc1n})-(\ref{abc3n}) one gets $k_1a^s=k_2b^s=k_3c^s$ 
for the irreversible cycle. 
Then, of course, we have
\begin{equation}
\left(\frac{\partial\sigma}{\partial a}\right)_s=0.
\label{mep1}
\end{equation}
Following a similar procedure, one easily obtains
\begin{equation}
\left(\frac{\partial\sigma}{\partial b}\right)_s=0=
\left(\frac{\partial\sigma}{\partial c}\right)_s.
\label{mep23}
\end{equation}
However, according to Eq.(\ref{eprir}), $\sigma(t)\ge 0$. 
So, the extremum at the steady state {\it is the 
minimum} and therefore, the steady state reached by the 
irreversible cycle is indeed the {\it state of equilibrium}.

\section{Equivalence of the approach with Onsager's original formulation}

There is an alternative way of defining the forces and 
the fluxes for a chemical reaction system than 
those given in Eqs (\ref{j1})-(\ref{x3}). 
Actually, this formalism was 
originally considered by Onsager \cite{ons1} 
to construct the EPR of a reversible triangular reaction network 
\cite{denb}. 
In this method, the flux is defined as the rate of 
change of concentration of a species and the 
corresponding force is the deviation of the chemical 
potential of that species from its equilibrium value.  
The fluxes are then written as 
\begin{equation}
J_a=\dot{a},
\label{j1n}
\end{equation}
\begin{equation}
J_b=\dot{b},
\label{j2n}
\end{equation}
\begin{equation}
J_c=\dot{c},
\label{j3n}
\end{equation}
with $\dot{a}+\dot{b}+\dot{c}=0.$
The corresponding forces are 
\begin{equation}
X_a=\mu_A^e-\mu_A(t)=T{\rm ln}\frac{a^e}{a(t)},
\label{x1n}
\end{equation}
\begin{equation}
X_b=\mu_B^e-\mu_B(t)=T{\rm ln}\frac{b^e}{b(t)},
\label{x2n}
\end{equation}
\begin{equation}
X_c=\mu_C^e-\mu_C(t)=T{\rm ln}\frac{c^e}{c(t)}.
\label{x3n}
\end{equation}
It is evident that defined this way, they 
are equally eligible to construct the expression of EPR 
in reversible as well as in irreversible chemical 
cycles. 
The EPR then becomes 
\begin{equation}
\sigma'(t)=\frac{1}{T}\sum_{i=a,b,c} J_i(t) X_i(t).
\label{eprn}
\end{equation}
As already mentioned, the concentrations of the species remain finite 
throughout the evolution of the reaction system with 
irreversible steps. 
Thus, a distinct advantage of Eq.(\ref{eprn}) is that 
the EPR {\it does not diverge} in the irreversible cycle. 
At equilibrium, the forces in Eqs (\ref{x1n})-(\ref{x3n}) 
(and the corresponding fluxes) vanish and 
and hence, the EPR becomes zero.

For the reversible cycle in Fig.\ref{fig1}, we can show the equality of 
Eq.(\ref{epr}) and Eq.(\ref{eprn}) straightforwardly. 
Starting from Eq.(\ref{eprn}) and using 
Eqs (\ref{j1n})-(\ref{x3n}) along with Eqs (\ref{abc1n})-(\ref{abc2n}) 
and $\dot{c}=-(\dot{a}+\dot{b})$, we obtain
$$T\sigma'(t)=\dot{a}(X_a-X_c)+\dot{b}(X_b-X_c)$$
$$=\dot{a}(\mu_C-\mu_A)+\dot{b}(\mu_C-\mu_B) $$
$$=(k_{-1}b(t)-k_1a(t))(\mu_B-\mu_A)+(k_{-2}c(t)-k_2b(t))(\mu_C-\mu_B)$$
$$+(k_{-3}a(t)-k_3c(t))(\mu_A-\mu_C)$$
\begin{equation}
=\sum_{i=1}^{3} J_i(t) X_i(t)=T\sigma(t).
\label{eql}
\end{equation} 
The only `condition' used in the above derivation 
is that of thermodynamic equilibrium, {\it i.e.,} Eq.(\ref{eqlbm}). 
To be also valid for the irreversible cycle in Fig.\ref{fig2}, 
the fluxes $J_i$ and the forces $X_i$ in Eq.(\ref{eql}) should be defined 
as given in Eqs (\ref{muca})-(\ref{j3ir}).  
This equivalence further strengthens the appoarch to 
formulate the EPR in an irreversible chemical cycle where 
both the forms can be used interchangeably. 
To convince ourselves further, 
in the next section, we will generalize the problem 
to an irreversible N-cycle by using fluxes and 
forces of the forms of Eqs (\ref{j1n})-(\ref{x3n}).

\section{EPR in an irreversible N-cycle}

\begin{figure}[tbh]
\centering
\rotatebox{270}{
\includegraphics[width=6cm,keepaspectratio]{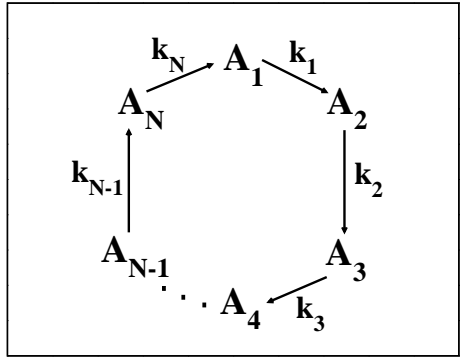}}
\caption{Schematic diagram of a cyclic irreversible reaction network 
containing $N$ number of species}
\label{fig3}
\end{figure}

In this section, we derive an expression of EPR 
in an irreversible cycle containing $N$ number of species, 
shown in Fig.\ref{fig3}, as a generelization of 
Eq.(\ref{eprn}). 
We denote the species by $A_i$ and their 
concentrations by $c_i(t), \, i=1,2,\cdots,N$. 
The rate equations are given as 
\begin{equation}
\dot{c_i}=-k_ic_i(t)+k_{i-1}c_{i-1}(t),\,i=1,2,\cdots,N
\label{req}
\end{equation}
with the boundary conditions $k_0=k_N$, $c_0(t)=c_N(t)$ and 
the constraint 
\begin{equation}
\sum_{i=1}^{N}\dot{c_i}=0.
\label{sumcdot}
\end{equation}
The fluxes are defined as 
\begin{equation}
J_i=\dot{c_i},\, i=1,2,\cdots,N
\label{fluxN}
\end{equation}
and the corresponding forces are 
\begin{equation}
X_i=(\mu_i^e-\mu_i(t)),\, i=1,2,\cdots,N.
\label{forceN}
\end{equation}
Then, using Eq.(\ref{sumcdot}) and the equality of 
chemical potentials of all the species at equilibrium, 
the EPR in the irreversible N-cycle can be written as 
\begin{equation}
\sigma'_N(t)=\frac{1}{T}\sum_{i=1}^{N} J_i X_i
=\frac{1}{T}\sum_{i=1}^{N-1} \dot{c_i}(\mu_N-\mu_i).
\label{eprN}
\end{equation}
Now at equilibrium with $\dot{c_i}=0,\,\forall i$, we have 
\begin{equation}
\frac{c_N^e}{c_i^e}=\frac{k_i}{k_N},\, i=1,2,\cdots,N-1.
\label{cratioN}
\end{equation}
Using Eq.(\ref{cratioN}), one can write 
\begin{equation}
\mu_N-\mu_i=\Delta \mu^{0}_{Ni}+T{\rm ln}\frac{c_N}{c_i}=
-T{\rm ln}\frac{c_N^e}{c_i^e}+T{\rm ln}\frac{c_N}{c_i}=
T{\rm ln}\frac{k_Nc_N}{k_ic_i}.
\label{muN}
\end{equation}
Then from Eqs (\ref{req}), (\ref{fluxN}) and (\ref{muN}), we 
can write Eq.(\ref{eprN}) as 
$$\sigma'_N(t)=
(k_Nc_N(t)-k_1c_1(t)){\rm ln}\frac{k_Nc_N(t)}{k_1c_1(t)}
+(k_1c_1(t)-k_2c_2(t)){\rm ln}\frac{k_Nc_N(t)}{k_2c_2(t)}
+\cdots$$
$$+(k_{N-3}c_{N-3}(t)-k_{N-2}c_{N-2}(t)){\rm ln}
\frac{k_Nc_N(t)}{k_{N-2}c_{N-2}(t)}
+(k_{N-2}c_{N-2}(t)-k_{N-1}c_{N-1}(t)){\rm ln}
\frac{k_Nc_N(t)}{k_{N-1}c_{N-1}(t)}$$
\begin{equation}
=k_1c_1(t){\rm ln}\frac{k_1c_1(t)}{k_2c_2(t)}+
k_2c_2(t){\rm ln}\frac{k_2c_2(t)}{k_3c_3(t)}+\cdots+
k_Nc_N(t){\rm ln}\frac{k_Nc_N(t)}{k_1c_1(t)}.
\label{eprN+}
\end{equation}
Eq.(\ref{eprN+}) is obviously the generalized form 
of Eq.(\ref{eprir}). 
Therfore, one can justify that $\sigma'_N(t)\ge 0$, 
following the similar procedure as outlined earlier. 
It is also evident that $\sigma'_N(t)=0$ corresponds to 
the state of equilibrium, 
with $\dot{c_i}=0,\,\forall i$. 

Now we test whether the MEPP holds for the irreversible N-cycle. 
From Eq.(\ref{eprN}), we can write
$$\left(\frac{\partial\sigma'_N}{\partial c_n}\right)=
\frac{\partial}{\partial c_n}\left(
\dot{c_n}{\rm ln}\frac{c_n^e}{c_n}+
\dot{c}_{n+1}{\rm ln}\frac{c_{n+1}^e}{c_{n+1}}\right)$$
\begin{equation}
=-k_n{\rm ln}\frac{c_n^e}{c_n}-\frac{\dot{c_n}}{c_n}
+k_n{\rm ln}\frac{c_{n+1}^e}{c_{n+1}}.
\label{eprmin}
\end{equation}
Then, at equilibrium, we get 
\begin{equation}
\left(\frac{\partial\sigma'_N}{\partial c_n}\right)_e=0,
\label{eprmep}
\end{equation}
as required by the MEPP.

Before concluding, we clarify that a linear irreversible nerwork 
cannot be treated using either formalism 
[Eq.(\ref{epr}) and Eq.(\ref{eprn})]. The reason is that, 
in the long-time limit, 
the amounts of all the species, except the terminal one, become 
zero, resulting in undefined chemical potentials.

\section{Conclusion}

In this work, we have discussed on the feasibility of an 
irreversible chemical cycle reaching equilibrium. 
The 
kinetic equations assert that the concentrations, and 
hence the chemical potentials, of all the constituents 
remain well-defined throughout the system evolution. 
This justifies a non-equilibrium thermodynamic study 
of such systems, where, in principle, nothing 
should diverge. The standard scheme fails 
in this respect, revealing a catastrophic behavior. 
We have established that, 
with appropriate redefinitions of forces, 
the divergence of EPR can be avoided. 
Using the formulation presented here, it is shown that any 
cyclic irreversible chemical network reaches a state 
of true thermodynamic equilibirum, indicated by zero EPR, though 
detailed balance is not satisfied. 
The equivalence of this 
approach with Onsager's originial formulation of EPR 
confirms its authenticity. 
For the reversible cycle, the proof is straightforward 
whereas, for the irreversible one, a modified 
description of force is required. 
The positivity of the expression of EPR  
in the irreversible cycle is established to show that the 
outcomes are thermodynamically consistent. 
Analysis of the MEPP further acknowledges that 
the EPR in the irreversible network 
has its minimum at the equilibrium sought. 
Therefore, we affirm that any irreversible chemical cycle does 
indeed reach and sustain equilibrium.

\section*{Acknowledgment}

K. Banerjee acknowledges the University Grants Commission (UGC), India 
for Dr. D. S. Kothari Fellowship.

\end{document}